# Rotationally Aligned Hexagonal Boron Nitride on Sapphire by High-Temperature Molecular Beam Epitaxy

Ryan Page[1], Yongjin Cho[2], Joseph Casamento[1], Sergei Rouvimov[3], Huili Grace Xing[1,2,4], and Debdeep Jena[1,2,4]

[1]*Department of Materials Science and Engineering, Cornell University, Ithaca, NY 14853, USA*
[2]*School of Electrical and Computer Engineering, Cornell University, Ithaca, NY 14853, USA*
[3]*Department of Electrical Engineering, University of Notre Dame, Notre Dame, IN 46556*
[4]*Kavli Institute at Cornell for Nanoscale Science, Cornell University, Ithaca, NY 14853, USA*
(Dated: Jul 17, 2019)

Hexagonal boron nitride (hBN) has been grown on sapphire substrates by ultrahigh-temperature molecular beam epitaxy (MBE). A wide range of substrate temperatures and boron fluxes have been explored, revealing that high crystalline quality hBN layers are grown at high substrate temperatures, >1600°C , and low boron fluxes, $\sim 1 \times 10^{-8}$ Torr beam equivalent pressure. In situ reflection high-energy electron diffraction revealed the growth of hBN layers with 60° rotational symmetry and the [11$\bar{2}$0] axis of hBN parallel to the [1$\bar{1}$00] axis of the sapphire substrate. Unlike the rough, polycrystalline films previously reported, atomic force microscopy and transmission electron microscopy characterization of these films demonstrate smooth, layered, few-nanometer hBN films on a nitridated sapphire substrate. This demonstration of high-quality hBN growth by MBE is a step toward its integration into existing epitaxial growth platforms, applications, and technologies.

## I. INTRODUCTION

Hexagonal boron nitride (hBN) is a two-dimensional wide bandgap semiconductor that has drawn intense interest in recent years due to its unique properties and wide range of applications. Individual sheets of hBN, like graphite, transition metal dichalcogenides, and other two-dimensional materials, are held together only by weak, interlayer, Van der Waals forces. The ability to isolate and stack layers of these different 2D materials has created the burgeoning field of Van der Waals heterostructures and devices [1-4]. As the only yet-known wide-bandgap 2D material, hBN will play a crucial role as an insulator, dielectric, substrate, or even potentially opto-electronic active region based on the ability to control its properties [1,2,5-7]. Additionally, hBN has been demonstrated as a promising material for ultraviolet light emission [8,9]. Despite the indirect character of its bandgap, hBN exhibits highly efficient radiative recombination through phonon-mediated transitions due to strong electron-phonon interactions [6,7,10-13]. It has also recently been shown to host point defects that function as bright, room temperature single photon sources, which can enable technologies such as quantum cryptography and precision sensing [14,15]. On top of these potential applications, the strong electron-phonon coupling and hyperbolic dispersion of hBN make it an exciting platform to study rich new physical phenomena such as phonon-polaritons and the effects of isotopes [12,16-20].

Currently, bulk crystal synthesis methods produce the highest quality hBN single crystals; millimeter scale crystals can be produced by high-pressure high-temperature (HPHT) or hBN precipitation methods [21,22]. Flakes exfoliated from these bulk crystals are employed in most demonstrations of hBN-based devices and applications. While this exfoliation manufacturing has enjoyed early successes in demonstrating high-quality Van der Waals devices and heterostructures, it is intrinsically non-scalable and highly time- and labor-intensive, precluding it from large-scale or commercial applications. To circumvent the limitations of bulk crystal exfoliation-based production, there is much interest in growing hBN with existing, mature thin film deposition techniques, such as chemical vapor deposition (CVD) or molecular beam epitaxy (MBE). Metalorganic chemical vapor deposition (MOCVD) has been able to produce mono- and few-layer flakes of hBN on metallic substrates with lateral grain dimensions on the order of tens of microns and has also demonstrated the growth of thick (~500 nm), hBN layers on c-plane sapphire substrates [6,23,24]. These MOCVD-grown films demonstrate excellent crystallinity and bright photoluminescence at 5.46 eV. However, due to the reactivity of the metalorganic precursors used, parasitic gas-phase reactions can be a concern for these processes. Recently, the growth of thin, layered hBN films on sapphire by high-temperature, low-pressure CVD has also been reported [25].

Work on the growth of hBN by MBE has been limited thus far, partially due to the high temperatures required for both the substrate and an elemental boron source. The recent availability of ultra-high temperature MBE substrate heaters and effusion cells has allowed these problems to be partially overcome. Several studies have previously demonstrated the growth of hBN films on both sapphire and highly-oriented pyrolytic graphite (HOPG) substrates [7,26]. They have reported high-quality mono-and few-layer hBN on HOPG. Few-layer growth of hBN by MBE has also been recently demonstrated on metallic substrates [27]. Work on the growth of hBN on c-plane

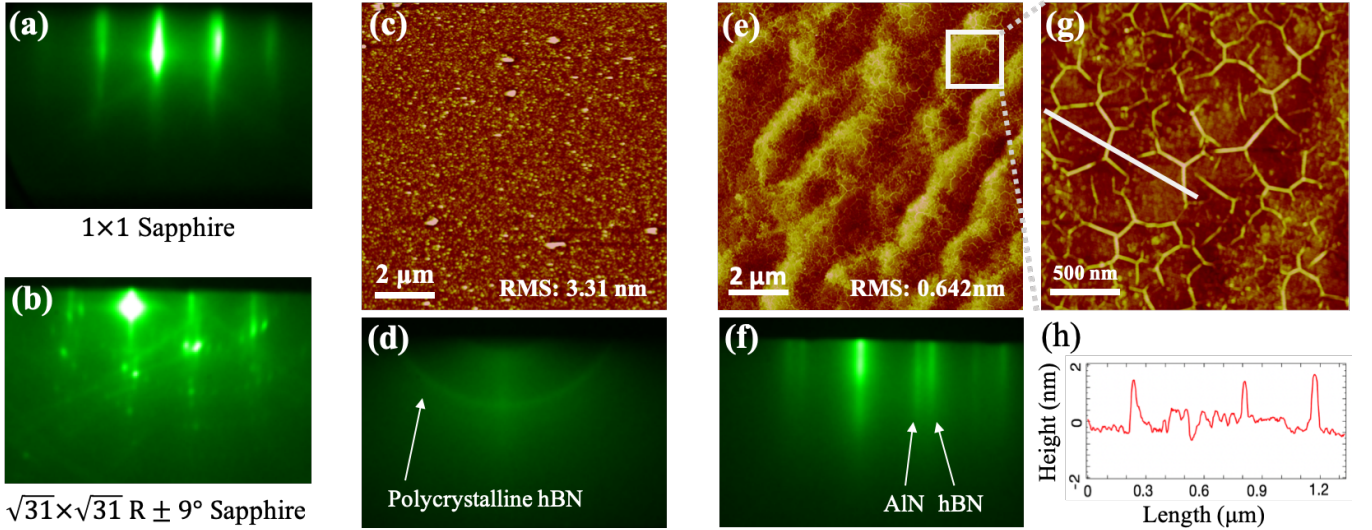

FIG. 1. (a), (b) RHEED patterns of the sapphire substrate taken on the same azimuth (a) before surface reconstruction (b) and after $\sqrt{31} \times \sqrt{31}$ R $\pm$ 9° surface reconstruction. (c), (d) Characteristic rough AFM morphology and RHEED pattern of polycrystalline hBN on sapphire grown at low substrate temperatures ($T_S < 1600$℃). (c) $10 \times 10\,\mu m$ AFM image and (d) the corresponding diffuse, ringed RHEED pattern demonstrate polycrystalline growth. (e)–(h) AFM and RHEED patterns of smooth, rotationally aligned hBN on sapphire grown at high substrate temperature ($T_S > 1700$℃) and low boron flux ($\sim 1 \times 10^{-8}$ Torr BEP): (e) $10 \times 10\,\mu m$ AFM image and (f) the corresponding RHEED pattern shows sharp streaks attributed to AlN and hBN (see text). (g) $2 \times 2\,\mu m$ AFM image of layered hBN on sapphire, exhibiting a dense network of wrinkles. (h) A line height profile of the white bar in (g) shows the height of the wrinkles to be 1–2 nm.

sapphire by MBE, however, yielded only rough, polycrystalline films.

Since many applications of hBN, from 2D FETs to hBN single photon sources require an insulating substrate, the growth of high-quality crystalline hBN on sapphire substrates is still highly desirable. Further, high quality hBN grown on sapphire allows a path towards integration with the established electronic and photonic device platforms of GaN and Al(In,Ga)N.

## II. EXPERIMENTAL

In this work, we report the growth and characterization of smooth, layered hBN on sapphire by ultra-high temperature MBE. All of the hBN films in this work were grown by high temperature molecular beam epitaxy (MBE) in a Veeco GENxplor system with substrate thermocouple temperatures up to 1800℃ Boron was provided by a high-temperature Knudsen effusion cell. For all growths, active nitrogen was supplied by a nitrogen plasma source operating at 200 W with 1.5 sccm gas flow. Prior to introduction to the MBE growth chamber, the sapphire substrates were ultrasonicated for 15 minutes in acetone and isopropanol, successively, and baked at 200℃ in UHV for 8 hours in the loadlock chamber.

It is known in MBE growth that the real surface temperature of the substrate deviates from the thermocouple reading, particularly at high temperatures. To calibrate and estimate the real substrate temperature, we used the well-studied surface reconstruction patterns of the sapphire substrate, as observed by *in situ* reflection high energy electron diffraction (RHEED)

[28]. We clearly observe the transition from the $1 \times 1$ surface to the $\sqrt{31} \times \sqrt{31}$ R $\pm$ 9° surface reconstruction pattern on the sapphire substrate, which occurs at a surface temperature of ~1250℃, at thermocouple readings varying from 1500℃ to 1600℃ [Fig. 1(a)].

Unlike the MBE growth of wurtzite III-nitrides, which demonstrate smooth, epitaxial growth under metal-rich flux ratios, optimal growth conditions for hBN are under an excess of active nitrogen. While a recent report suggests an influence of active N flux on hBN morphology on HOPG, the rate-limiting element, boron, will govern the overall morphology and growth rate of these films [29]. Therefore, in this work, the nitrogen flux was kept constant while the boron flux and the substrate temperature were varied.

Initial growths consistently exhibited the Raman $E_{2g}$ peak and absorption peak characteristic of hexagonal phase boron nitride, but had a rough, polycrystalline morphology, similar to previous reports [7]. However, as the growth parameter space of boron flux and substrate temperature was spanned, we observed that the resultant hBN films on sapphire fell consistently into one of two growth modes: rough, highly polycrystalline films, consistent with the results previously published in Ref. 7, or a heretofore unreported growth mode resulting in smooth, layered, and rotationally aligned two-dimensional hBN layers on the three-dimensional sapphire substrate.

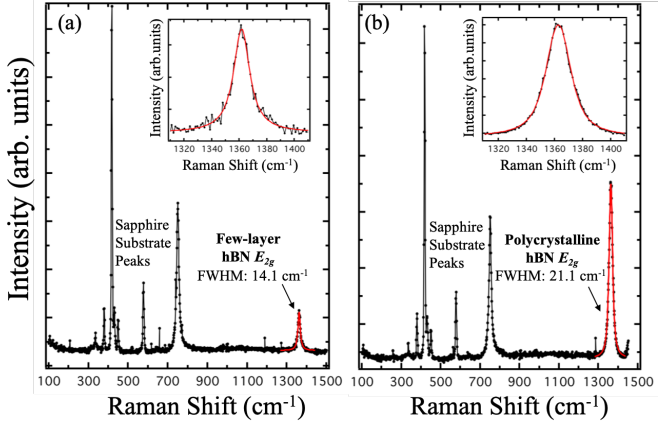

FIG. 2. Raman Spectra of (a) few-layer hBN and (b) polycrystalline hBN on sapphire substrates. The Raman-active $E_{2g}$ mode in each spectrum is fitted to a Voigt function (red line). The $E_{2g}$ peak in polycrystalline hBN is broader and more intense, relative to the sapphire substrate peaks, implying a thicker film of lower crystalline quality. The few-layer hBN demonstrates a less intense, but much narrower peak, suggesting a thinner film of higher crystalline quality.

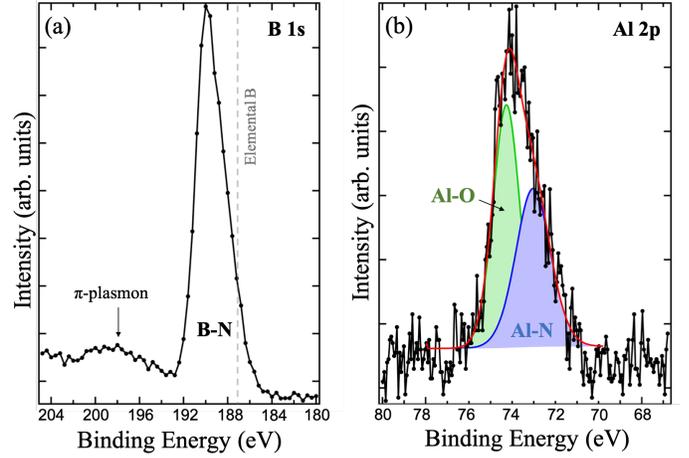

FIG. 3. XPS spectra of (a) the B 1s and (b) Al 2p peaks of few-layer hBN on sapphire. The position of the B 1s peak centered at 190.2 eV, is consistent with B-N bonding. A less intense peak is also observed 9 eV to higher energy, attributed to a $\pi$ plasmon resonance, a fingerprint of $sp^2$ bonding. The Al 2p peak can be deconvoluted into the sum of two Gaussian peaks centered at 74.2 eV and 73.1 eV, attributed to Al-O and Al-N bonding, respectively, demonstrating that the sapphire substrate was nitridated during the initial growth.

The growth of hBN by MBE depends critically on the substrate temperature, $T_S$. Below a thermocouple temperature of approximately 1600℃, the hBN films are consistently rough and polycrystalline across all boron flux conditions used. The polycrystallinity of these films is indicated by a diffuse, ringed RHEED pattern, which correlates with a rough surface morphology observed in atomic force microscopy (AFM) [Figs. 1(c)-1(d)]. As the substrate temperature is increased above this threshold and up to $T_S = 1800℃$ (the limit of the substrate heater), the surface morphology depends on the impinging boron flux. At high temperatures (1650-1800℃) and high boron fluxes, above approximately $1.5 \times 10^{-8}$ Torr BEP (a boron effusion cell temperature, $T_B$, of 1950℃), the films are once again polycrystalline and rough, with the RHEED pattern proceeding within minutes from the reconstructed sapphire surface pattern to a diffuse pair of concentric rings. This is markedly different from the RHEED pattern observed under the conditions of high substrate temperatures and low boron fluxes of approximately $0.6 - 1.0 \times 10^{-8}$ Torr BEP ($T_B = 1850 - 1900℃$), which is characterized by bright, sharp streaks with 60° rotational symmetry, suggestive of a rotationally oriented, highly crystalline film of hBN [Fig. 1(f)].

Further, films grown under these high temperature, low boron flux conditions exhibit a dramatically different AFM surface morphology from the polycrystalline films, as can be seen in Fig. 1. In the $10 \times 10$ $\mu$m AFM image in Fig. 1(e), the step bunch terraces of the underlying sapphire substrate are visible, but are overlaid with a dense network of fine ripple-like features. These features are more clearly observed in the $2 \times 2$ $\mu$m image [Fig. 1(g)], which shows them intersecting preferentially at 120° angles. The height of these features is 1-

2 nm, as seen in the AFM line height profile plot in Fig 1(h). The height and density of these ripples are identical to those observed in transferred hBN flakes [30]. This effect is due to the negative thermal expansion normal to the c-plane observed in two dimensional materials, caused by the increased population of vibrational modes along the c-axis at high temperatures.

Raman spectroscopy was performed on all samples to confirm the presence of hexagonal phase boron nitride. A 532 nm laser confocal Raman microscope equipped with an 1800 mm$^{-1}$ diffraction grating was used for all measurements. Fig. 2 shows the Raman spectra of hBN films characteristic of each growth mode; each film was grown for eight hours under the same boron flux of $1.0 \times 10^{-8}$ Torr BEP ($T_B = 1900℃$,),) but two different substrate thermocouple temperatures. Fig. 2(a) was taken from a film grown at $T_S = 1750℃$ while Fig. 2(b) was from a film grown at $T_S = 1650℃$. Visible in both spectra is the peak corresponding to the $E_{2g}$ Raman-active vibrational mode of hBN. The peaks are fitted with a Voigt function (red line). The spectral position of the $E_{2g}$ peak (~1363 cm$^{-1}$ in our spectrometer) is consistent.

Comparing the two spectra, however, gives information about the relative film quality and thickness of the two samples. Films grown in the smooth growth mode consistently exhibited sharper, less intense peaks, while the rough, polycrystalline films exhibited relatively broader and more intense peaks. the peak intensities, relative to the sapphire substrate Raman peaks, suggest that the lower substrate temperature yields a thicker film. This qualitative observation is confirmed with subsequent AFM and TEM characterization. This temperature dependence of film thickness, which has been previously documented, implies that the film growth rate is governed by competing processes of crystal growth and decomposition or reduced sticking coefficients at high temperatures [7].

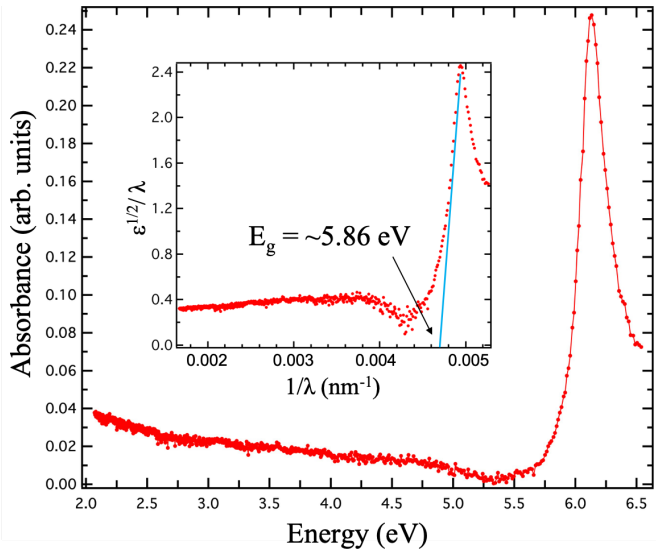

FIG. 4. Absorbance spectrum of few-layer hBN shows the characteristic peak centered at ~6.1 eV. A Tauc plot (inset) was used to estimate the band gap of the material. The Tauc method for an indirect band gap semiconductor was used to estimate the optical band gap to be ~5.86 eV, assuming a three-dimensional density of states.

The chemical composition and bonding character of these films were studied with *ex situ* X-ray photoemission spectroscopy (XPS). Samples were analyzed with a Surface Science Instruments SSX-100. Monochromatic Al Kα x-rays (1486.6 eV) with a 1 mm diameter beam size were used and photoelectrons were collected at a 55° emission angle. A hemispherical analyzer determined electron kinetic energy using a pass energy of 150 V for survey and high sensitivity scans and 50 V for high resolution scans. A constant energy flood gun was used for charge neutralization for insulating hBN samples. For the crystalline hBN grown at high temperature on sapphire, The B 1s core level peak was identified at 190.2 eV [Fig. 3(a)], which is consistent with prior reports of B-N bonding. Additionally, the B 1s XPS peak exhibits a small, broad peak shifted 9 eV to higher energy. This peak has been reported in hexagonal phase boron nitride, attributed to a π plasmon resonance and is a fingerprint of $sp^2$ bonding in boron nitride [31,32].

Due to the ultra-high growth temperatures and slow growth rate, which is estimated by transmission electron microscopy later in the text to be approximately one monolayer per hour under these conditions, surface nitridation of the $Al_2O_3$ substrate during early stages of growth is possible [33,34]. To investigate this, we analyzed the Al 2p peak with high resolution XPS [Fig. 3(b)]. The peak exhibited asymmetric broadening, indicative of a mixture of chemical bonding environments, and was deconvoluted into two contributions, an Al-O bonding environment peak centered around 74.2 eV, and an Al-N bonding environment peak centered around 73.1 eV [35,36]. This indicates that during the initial stage of growth, the sapphire surface is partially converted into aluminum nitride (AlN). The presence of a small amount of AlN, which has a direct bandgap with a similar energy to that of the indirect bandgap of hBN, could convolute the optical properties of these films (e.g. photoluminescence spectra), which will be investigated in the future.

To isolate and characterize the sapphire nitridation, a sapphire substrate was loaded into the growth chamber and subjected to the same temperature and active nitrogen flux conditions used for the high quality hBN layers, but without exposure to boron. During the 8-h nitridation, a small change in the RHEED pattern was observed and the presence of Al-N bonding was confirmed by XPS. However, the AFM surface morphology of this sample showed step-bunched terraces similar to the sapphire substrate. Additionally, to begin to better understand the growth mechanisms of these films, XPS was performed on a rough, polycrystalline sample grown under unoptimized conditions. XPS again demonstrated Al-N bonding character, indicating that this nitridation layer is not itself sufficient for the growth of high crystalline quality films.

Figure 4 shows the characteristic hBN absorbance spectrum of a crystalline hBN film grown on sapphire. Due to the *q* dependence of the exciton-phonon interaction matrix elements and the location of the band extrema, which are not at the zone-center, the absorbance spectrum of hBN exhibits a sharp peak at the band edge, rather than the usual, sustained absorption of photons with energy above the bandgap of a material [13]. The bandgap can be estimated using the method of Tauc, given by the following equation and the inset plot in Fig. 4 [37]:

$$\omega^2 \varepsilon \sim (\hbar\omega - E_g)^2$$

where $\omega$ is the frequency of the light, $\varepsilon$ is the absorbance, $\hbar$ is the reduced Planck constant, and $E_g$ is the optical bandgap of the material. hBN has an indirect bandgap, which sets the value of the exponent as 2 on the absorbance in the y-axis of the inset of Fig. 4. Additionally, it was assumed that this material has a three-dimensional density of states, since the films studied here consist of multiple layers of hBN. The extracted value of the bandgap is very similar to the values estimated by similar methods in the literature and is close to the experimental value of the indirect bandgap of hBN, 5.9 eV [38].

A quantitative analysis of the RHEED patterns, specifically the pairs of streaks observed in the RHEED image in Fig. 1(f) can yield information about the reciprocal lattices of the material sampled by the RHEED electron beam. The known lattice parameters of the sapphire substrate were used as a calibration to convert from CCD pixel spacing to a length in reciprocal space. The calibration images were taken at a thermocouple temperature of 1600 ℃, shortly before the sapphire underwent surface reconstruction. The two sets of streaks visible in the in the RHEED image in Fig. 1(f), combined with the presence of Al-N bonding character in the XPS spectra allow us to attribute the inner and outer streaks to AlN and hBN, respectively. Using the sapphire substrate as a reference, the $a$-lattice parameters of AlN and hBN were estimated by RHEED to be ~2.9 and ~2.4 Å. This estimate is slightly smaller than the known values for both materials, however the ratio of the two values matches closely that of the known values, implying a slight error in the sapphire calibration, likely caused by a slight misalignment between the RHEED source and the high-symmetry axes of the substrate. Comparing the azimuths of the high-symmetry directions indicates that the hBN and AlN layers are rotated 30° with respect to the sapphire substrate. This rotation has been reported in the growth of AlN layers on sapphire, as well as the MOCVD growth of hBN on sapphire, in order to minimize lattice mismatch [25,32]. Rotational alignment has also been analyzed for hBN grown on polycrystalline copper substrates; it was shown that the hBN edges preferentially align to a close-packed direction of the underlying substrate surface, with flakes nucleated on the hexagonally-symmetric (111) planes exhibiting the greatest degree of alignment [24].

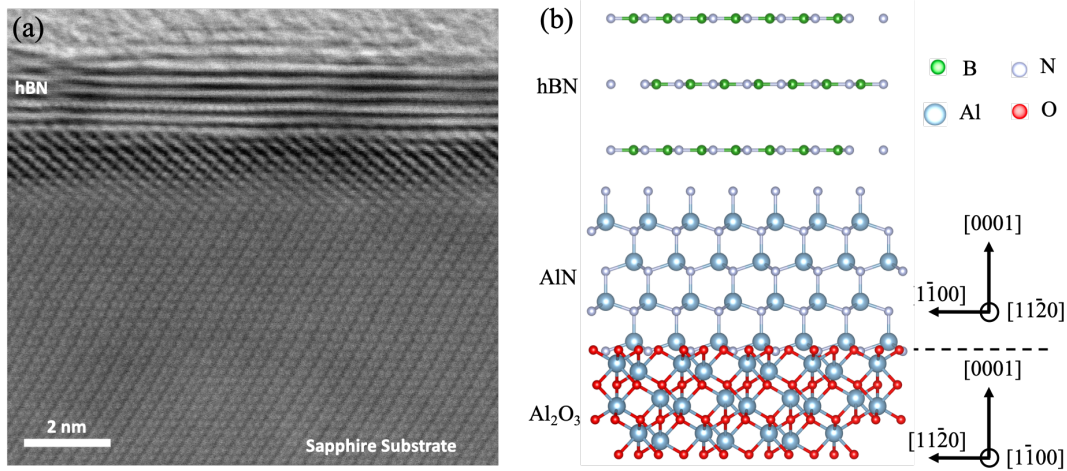

FIG. 5. (a) Cross-sectional STEM image of hBN grown on sapphire by MBE shows smooth, layered growth of few-layer films. Evidence of surface nitridation is directly corroborated by the presence of an interfacial layer between the bulk sapphire substrate and the hBN layers with a crystal structure consistent with [11$\bar{2}$0] AlN. (b) A proposed schematic of this sapphire/AlN/hBN heterostructure, demonstrating the 30° rotation fo the AlN with respect to the sapphire. The Al-polarity of the AlN shown in the schematic is not experimentally verified.

To further elucidate the structural properties of the hBN films and to directly observe the thickness and presence of sapphire surface nitridation, high-resolution cross-sectional transmission electron microscopy (TEM) and scanning TEM (STEM) were performed on a smooth hBN film. Fig 5(a) shows a bright-field STEM cross section of an hBN film on a sapphire substrate, demonstrating the layered hBN film on the substrate. While the underlying layers are fully coalesced, the top layer does exhibit discontinuity and island formation, resulting in a non-uniform integral number of layers across the film. Image analysis of the hBN layers yields an interlayer spacing of approximately 3.2-3.4 Å, consistent with the equilibrium spacing of hBN. The thin, high contrast region observed near the sapphire-hBN interface in Fig. 5(a) and the accompanying change in crystal structure is attributed to AlN. The structure and relative orientation of the $Al_2O_3$-AlN-hBN heterointerface is schematically shown in Fig. 5(b) [39]. It should be noted that the Al-polarity of the AlN layer is not experimentally confirmed. The thickness of the film, which was grown for eight hours, also confirms the extremely slow growth rate of highly crystalline and rotationally aligned hBN under these growth conditions: roughly one monolayer per hour.

## III. CONCLUSIONS

In conclusion, we have demonstrated the growth of smooth layers of two-dimensional hexagonal boron nitride on a three-dimensional sapphire substrate. Mapping the MBE growth parameter space of hBN on sapphire has revealed that extremely high substrate temperatures and low boron fluxes are necessary for growth of high-quality films, possibly by minimizing crystal nucleation events or effecting competitive crystal growth and decomposition, either of which would result in the observed slow net film growth rate. This work is an important step toward the integration of epitaxially-grown hBN layers directly into devices and applications for DUV lighting, single photon emission, and many other fields.

## ACKNOWLEDGEMENTS

This work was supported in part by an AFOSR (Grant No. FA9550-17-1-0048) award monitored by K. Goretta, the NSF Award No. DMREF-1534303, the NSF RAISE-TAQS Award No. 1839196, the NSF EFRI-2DARE Award No. EFRI 1433490, and by the Cornell Center for Materials Research with funding from the NSF MRSEC Program No. DMR-1719875. This work was performed in part at the Cornell NanoScale Science and Technology Facility (CNF), a member of the National Nanotechnology Coordinated Infrastructure (NNCI), which is supported by the National Science Foundation (Grant No. NNCI-1542081).